\documentclass[twocolumn, trackchanges]{aastex631}
\usepackage{amsmath,amsfonts}
\usepackage{xspace}

\def\lsim{\mathrel{\raise.3ex\hbox{$<$\kern-.75em\lower1ex\hbox{$\sim$}}}}
\def\gsim{\mathrel{\raise.3ex\hbox{$>$\kern-.75em\lower1ex\hbox{$\sim$}}}}
\def\gtwid{\mathrel{\raise.3ex\hbox{$>$\kern-.75em\lower1ex\hbox{$\sim$}}}}
\def\proptwid{\mathrel{\raise.3ex\hbox{$\propto$\kern-.75em\lower1ex\hbox{$\sim$}}}}

\def\m87{M87$^*$\xspace}

\usepackage{url}

\graphicspath{{./Plots/}}
\usepackage{autobreak}

\begin{document}

\title{The Role of Interferometric Phase in Measuring Black Hole Photon Rings}

\author[0009-0004-6333-1744]{Sol Gutiérrez-Lara}
\affiliation{Black Hole Initiative at Harvard University, 20 Garden Street, Cambridge, MA 02138, USA}
\affiliation{Center for Astrophysics $|$ Harvard \& Smithsonian, 60 Garden Street, Cambridge, MA 02138, USA}
\affiliation{%
Department of Physics, Jefferson Laboratory, Harvard University, Cambridge, MA 02138, USA
}%

\author[0000-0002-7179-3816]{Daniel~C.~M.~Palumbo}
\affiliation{Black Hole Initiative at Harvard University, 20 Garden Street, Cambridge, MA 02138, USA}
\affiliation{Center for Astrophysics $|$ Harvard \& Smithsonian, 60 Garden Street, Cambridge, MA 02138, USA}

\author[0000-0002-4120-3029]{Michael D. Johnson}
\affiliation{Black Hole Initiative at Harvard University, 20 Garden Street, Cambridge, MA 02138, USA}
\affiliation{Center for Astrophysics $|$ Harvard \& Smithsonian, 60 Garden Street, Cambridge, MA 02138, USA}

\begin{abstract}
The Event Horizon Telescope (EHT) captured the first images of a black hole using Very Long Baseline Interferometry (VLBI). In the near future, extensions of the EHT such as the Black Hole Explorer (BHEX) will allow access to finer-scale features, such as a black hole's ``photon ring.'' In the Kerr spacetime, this image structure arises from strong gravitational lensing near the black hole that results in a series of increasingly demagnified images of each emitting region that exponentially converge to a limiting critical curve. Exotic black hole alternatives, such as wormholes, can introduce additional photon rings. Hence, precisely characterizing multi-ring images is a promising pathway for measuring black hole parameters, such as spin, as well as exploring non-Kerr spacetimes. Here, we examine the interferometric response of multi-ring systems using a series of 1) simple geometric toy models, 2) synthetic BHEX and EHT observations of geometric models, and 3) semi-analytic accretion models with ray-tracing in the Kerr spacetime. We find that interferometric amplitude is more sensitive to the shape of the photon ring, while interferometric phase is more sensitive to its displacement, which is most sensitive to black hole spin. We find that for models similar to Messier~87* (M87*), the relative displacement of the first strongly lensed image from the weakly lensed direct image is approximately $1\,\mu {\rm as}$ per unit dimensionless spin, yielding an expected phase signature on a 25\,G$\lambda$ baseline of $\sim44^\circ$ per unit spin. 
\end{abstract}

\section{Introduction} \label{sec:intro}

% Black hole imaging:
The Event Horizon Telescope (EHT) recently produced the first images of a black hole using the technique of Very Long Baseline Interferometry (VLBI) at 1.3\,mm wavelength \citep{EHT_M87_I,EHT_SGRA_I}. In the near future, extensions of the EHT such as the Black Hole Explorer \citep[BHEX;][]{Johnson_2024,Marrone_2024,Lupsasca_2024} will allow access to finer-scale features, such as a black hole's ``photon ring'' \citep{Johnson_2020}. 

\begin{figure*}[t]
    \centering
\includegraphics[width=0.9\textwidth]{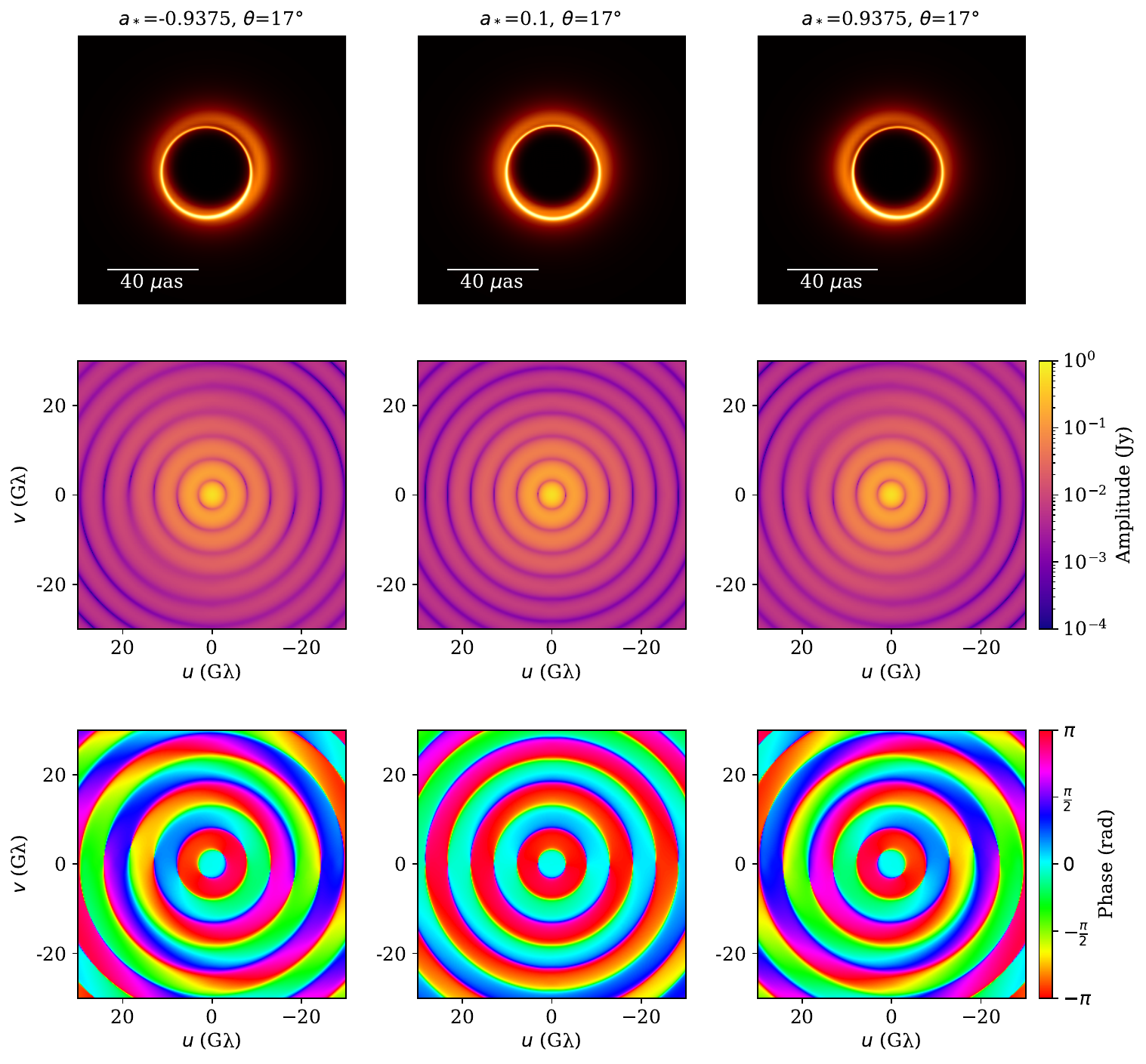}
    \caption{Three images of spinning black holes and their associated interferometric amplitude and phase responses. Top: ray-traced images of black holes of high spin oriented into the page (left), low spin oriented out of the page (middle), and high spin oriented out of the page (right). To isolate the effects of spin, we have fixed the viewing inclination at $\theta=17^\circ$, and we assume isotropic emission from a plasma at rest with respect to the local zero angular momentum observer (ZAMO). Middle: the amplitude response for each image, sampled after centroiding the image}. Bottom: the phase response for each image. The spin has a subtle effect on visibility amplitudes and a pronounced effect on visibility phases.
    \label{bhspin}
\end{figure*}

% Photon ring
A photon ring is a bright, thin ring of emission on the image seen by a distant observer around a source of gravity so strong that it can pull photons into partial or full orbits. Each additional orbit requires exponentially tighter constraints on a photon's trajectory --- converging on unstable photon orbits around the compact object \citep{Bardeen_1973,Teo_2003} --- producing exponentially sharper rings. Hence, the photon ring on the image of a black hole is composed of a nested series of increasingly sharp subrings formed by photons that have orbited the black hole multiple times \citep{Johnson_2020}. The subrings are commonly numbered as follows: $n=0$ refers to the most weakly-lensed ``direct'' image of light that reaches the observer, $n=1$ refers to the ring formed by photons that completed a half-orbit before escaping, $n=2$ corresponds to a full orbit, increasing by half-orbits with each $n$.
The features of the photon ring and their asymptotic critical curve encode information about an object's surrounding spacetime and gravitational structure \citep[see, e.g.,][]{Takahashi_2004,Johannsen_2010, Lupsasca_2020}. In addition to black holes, other exotic compact objects exhibit similar photon ring structure \citep[see, e.g.,][]{Wielgus_2020}.

% Black holes spin
A property of particular interest is the black hole's spin. Spin drags and distorts the surrounding spacetime, significantly affecting both photon orbits and the photon ring.
%which has drastic effects on the appearance of nearby emission–– including the photon ring \citep{Gralla_2019}. 
For example, the presence of spin produces a shift in the center of the critical curve and flattens it slightly on one side \citep[see, e.g.,][]{Takahashi_2004,Johannsen_2013,Farah_2020}. These effects create an opportunity for direct measurements of spin through precise measurements of the photon ring.

%Interferometry
Interferometers such as the EHT and BHEX measure the source morphology in the Fourier domain, sampling complex-valued ``visibilities.'' Thus, to evaluate the prospects for spin measurements with an interferometer it is imperative to characterize how the spin affects the Fourier components of an image. 
While most previous studies have focused on the \textit{amplitude} of the visibility response, the \textit{phase} of the visibility response is equally important and is sensitive to some effects (e.g., an overall image displacement) that do not affect visibility amplitudes.

To motivate this study, \autoref{bhspin} shows three example black hole models with varying spin (denoted hereafter as ``$a_*$") and their interferometric phase responses (hereafter, simply ``phase''). The black hole models were generated using \texttt{KerrBAM}, a semi-analytic ray-tracing code for modeling optically thin synchrotron or isotropic emission of an accretion disk, which can hold plasma properties fixed while varying viewing angle and spin \citep{Palumbo_2022}. 
In the image domain, the shift of the $n=1$ relative to the $n=0$ subring across spin values is obvious by eye, while its stretch is hardly perceptible. The visibility amplitudes remain nearly constant across spin values, while the phase response varies significantly. At low spin, it closely resembles that of a Bessel function of zeroth order with a phase that jumps discontinuously between $0$ and $\pi$ radians, excepting small deviations due to the nonzero inclination of the system which we return to in \autoref{sec:nonspacetime}. At high spin the single-valued concentric circular slices of the Bessel function blur into continuous gradients with a slight spiral, reflecting the fact that the visibility becomes complex-valued and the ring is slightly stretched.

\autoref{bhspin} suggests that the effects of spin on the photon ring are naturally encoded in the complex interferometric phase. We investigate this question by isolating such effects and studying their phase responses using both analytic and semi-analytic methods. In \autoref{sec:analytic}, we analytically study the interferometric response using a geometric ring model that approximates the relevant spacetime effects of spin. In \autoref{sec:observability}, we evaluate the prospects for measuring the interferometric features for multi-ring systems for the EHT and BHEX. In \autoref{sec:nonspacetime}, we assess how aspects of the emission such as the plasma properties and viewing inclination will affect the predicted interferometric response. We summarize our conclusions in \autoref{sec:conclusion}.

\begin{figure*}[ht]
    \centering
\includegraphics[width=0.9\textwidth]{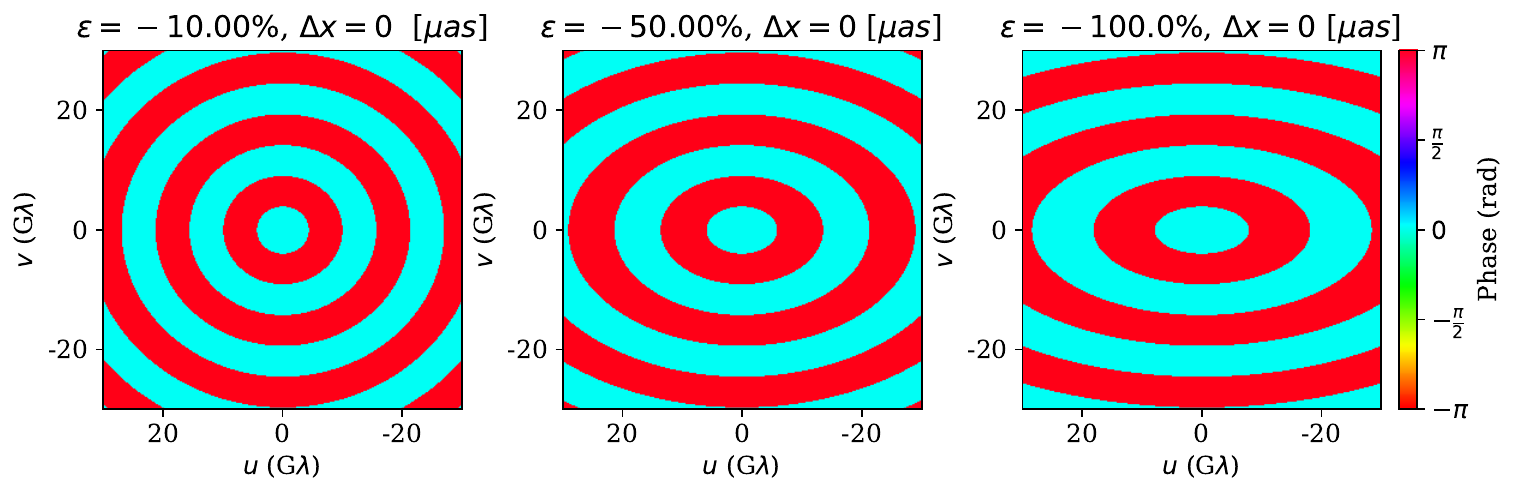}
\includegraphics[width=0.9\textwidth]{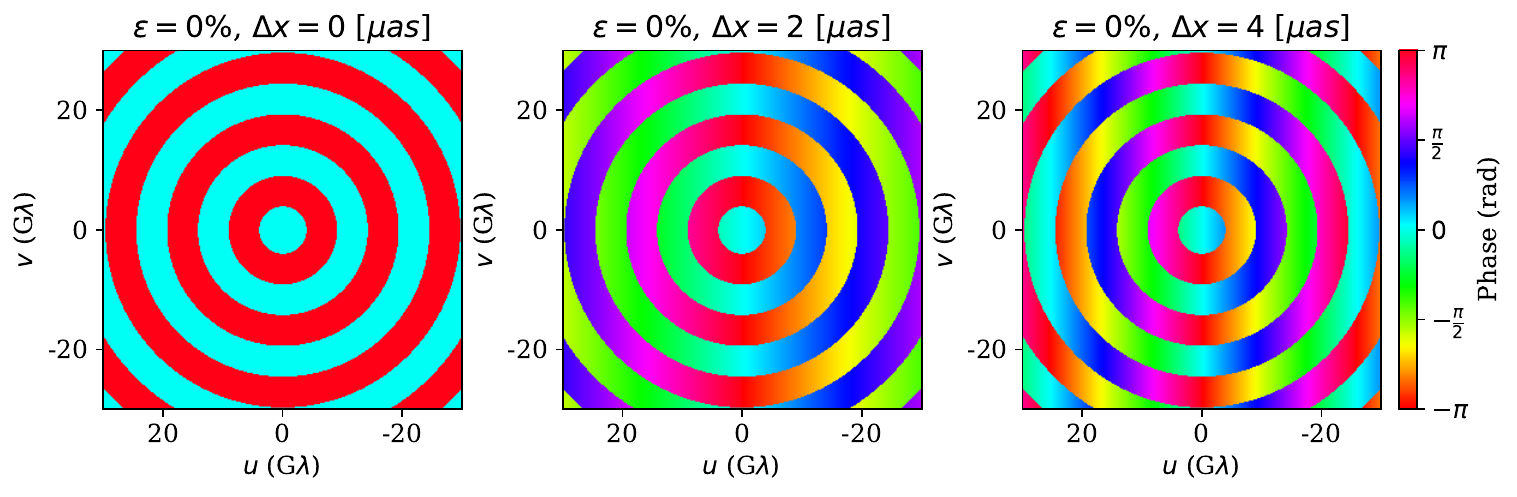}
\includegraphics[width=0.9\textwidth]{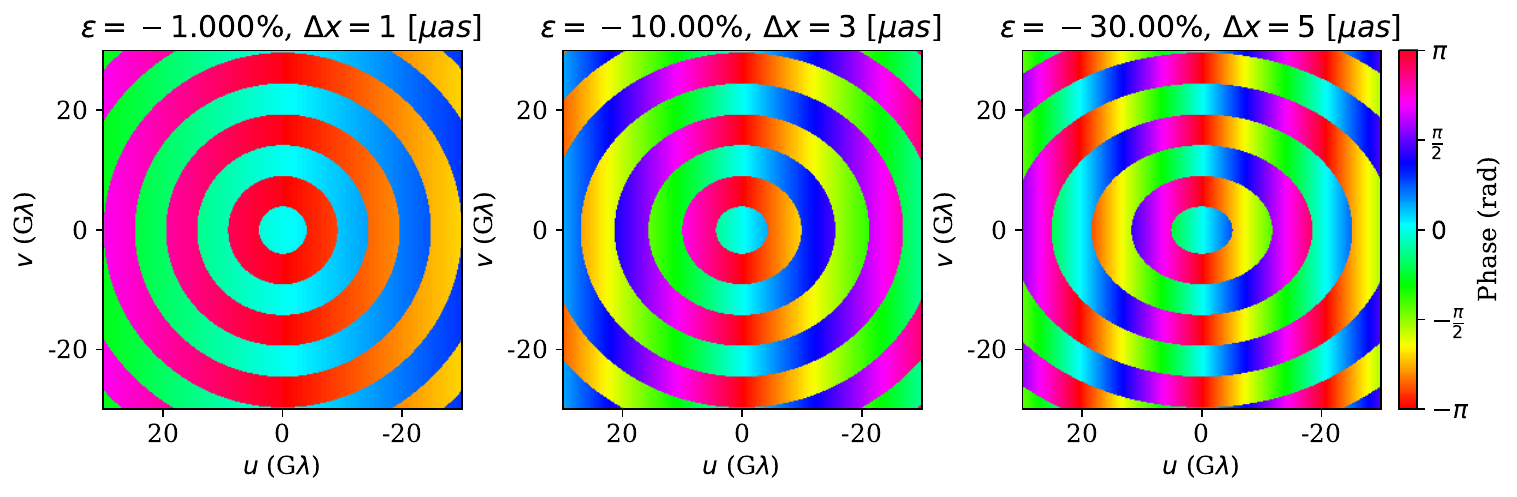}
    \caption{
    Interferometric phase response for a single geometric ring, varying its degree of stretching $\epsilon$ (row 1), the displacement of its center $\Delta x$ (row 2), and the combined effects of stretching and shifting (row 3). 
    Vertical stretching in the image corresponds to horizontal stretching in the interferometric response. 
    Shifting in the image introduces a phase slope that grows linearly with the baseline projected along the shift direction, thus replacing the monochromatic slices with a phase that varies smoothly and periodically. For this single-ring case, both shifting and stretching preserve the discontinuous phase jumps between slices. 
    }
    \label{onering}
\end{figure*}

\section{Analytic Characterization of the Photon Ring}
\label{sec:analytic}

In this section we consider analytic models of the direct and indirect images. 
We simplify the lensing problem by beginning with a geometric construction, following the treatment in \citet{Johnson_2020}. That is, we assume that each subring can be described by an infinitesimally thin Dirac delta ring (hereafter a ``$\delta$-ring'') convolved with a Gaussian kernel. We then analyze the visibility response of stacked rings under the expected spacetime distortions from spin modeled as simple geometric mutations -- stretching the ring and shifting its center -- both of which have simple interferometric responses. Since the Fourier decomposition of two superimposed objects in the image domain is simply the sum of their respective Fourier responses, it is then straightforward to extend this analytic model of a single ring to a system with multiple superimposed rings. 

The purpose of this section is to analytically isolate the spin effects on ring morphology. We focus on azimuthally symmetric rings to study shifting and stretching, both of which preserve angular structure. Given that the detailed azimuthal structure of images can depend on both spacetime and astrophysical effects, in \autoref{sec:nonspacetime} we consider rings formed by black holes with additional astrophysical properties and disk inclination, which can introduce more complicated distortions such as brightness asymmetries.

\subsection{A Single Ring}
\label{sec:onering}

Consider a $\delta$-ring with unit flux density, intensity $I(x,y)$, and radius $R$ centered at the image origin, described by \autoref{eq:im}.
The visibility response (denoted ``$V(u,v)$" with baseline coordinates $u$ and $v$) of this source distribution is a Bessel function $J$ of zeroth order, with a phase that jumps back and forth between $0$ and $\pi$ radians as the baseline length increases:
\begin{align}
    I(x,y) = \frac{1}{2 \pi R}  \delta(\sqrt{x^2+y^2}-R); \label{eq:im} \\
    V(u,v) = J_0\left(2 \pi R \sqrt{u^2+v^2}\right).
\end{align}
To integrate the primary effects of spin, we shift the photon ring off-center, and stretch it into an ellipse. We also add a nonzero thickness to the ring by convolving it with a Gaussian in the image domain, equivalently multiplying by a Gaussian in the Fourier domain. The effects combine simply in the interferometric response:
\begin{align}
V(u,v) &= e^{-2\pi^2 \sigma^2 (A^2u^2+B^2v^2)} \\
\nonumber   \qquad {} &\times J_0\left(2 \pi R \sqrt{A^2u^2+B^2v^2}\right)
e^{-i2\pi(u\Delta x + v\Delta y)}.
\end{align}

%squish -> stretch
Here, $\sigma$ is the standard deviation of the Gaussian blurring kernel, $A$ and $B$ are the horizontal and vertical stretch factors of the ring image respectively, and in the $J_0$ argument we have replaced its angular radius $R$ with the stretched radius. 
Additionally, $\Delta x$ and $\Delta y$ are the horizontal and vertical displacements respectively of the ring image center. They appear in the Fourier domain as an overall phase factor, growing linearly with the baseline.
We quantify the degree of stretching with $\epsilon$, which we define to be:
\begin{align}
    \epsilon = 1 - \frac{B}{A}.
\end{align}
A ring that is stretched vertically has $\epsilon<0$. For simplicity we set $B=1$ for every model in \autoref{sec:analytic}, horizontally compressing the ring with $A<1$.
Finally, we denote the interferometric phase by $\psi \equiv {\rm arg}\left(V \right)$.
\autoref{onering} shows the effects of a variety of such distortions. While the complex visibility is continuous, the discontinuities in the wrapped phase across $\psi=\pm \pi$ are detectable signatures that persist in large swings in the observed closure phases from rings, as seen in EHT observations of M87* in \citet{eht3_m87} and \citet{eht4_m87}.

\begin{figure*}[t]
    \centering
    \includegraphics[width=0.9\textwidth]{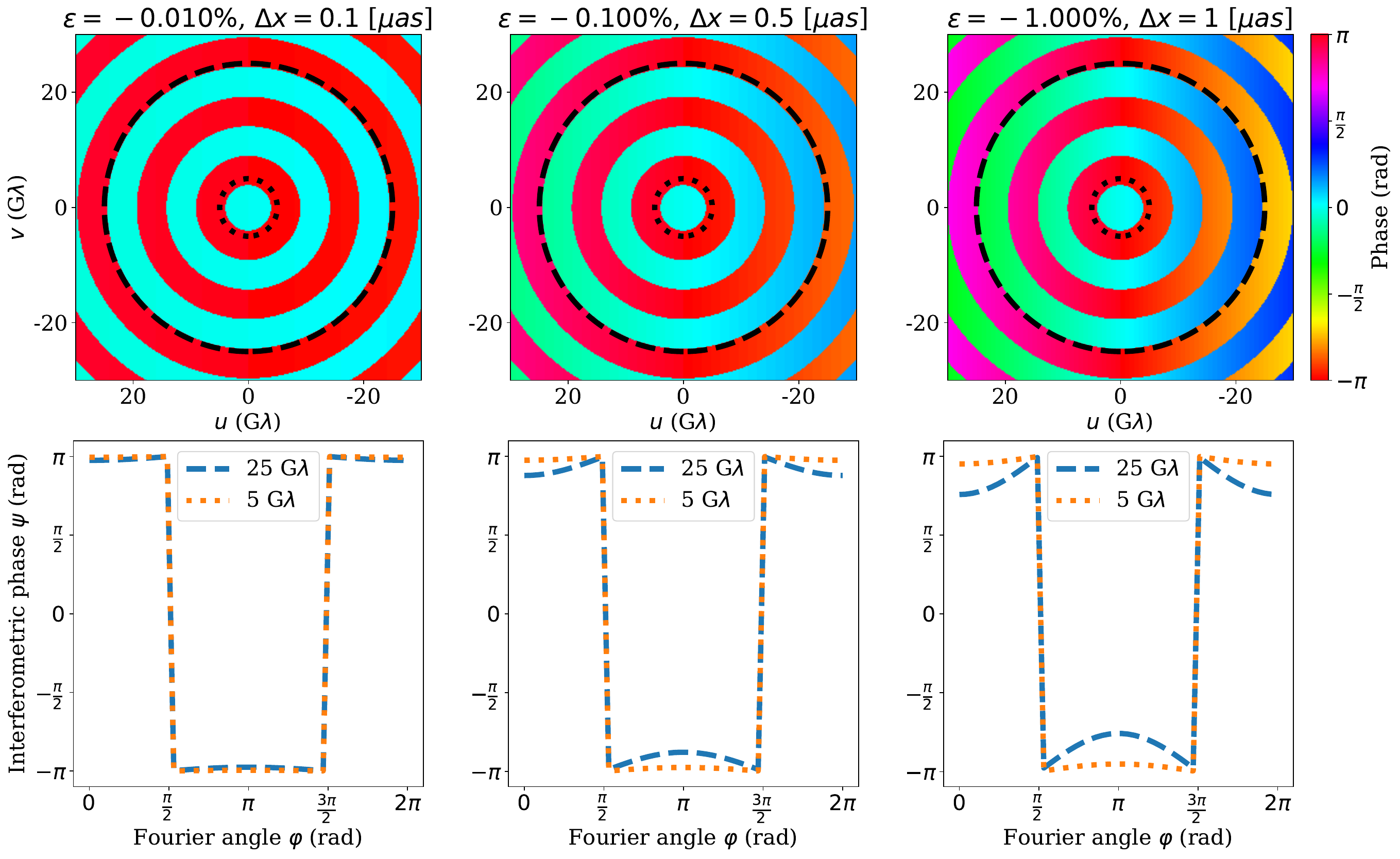}
\caption{Row 1: visibility phases from a single-ring model with different parameters. Row 2: corresponding azimuthal variation (along the Fourier angle $\varphi$) of the phase $\psi$ at two different fixed baseline lengths, chosen to match typical Earth-Earth and Earth-space baselines.
The marked circles in row 1 show the fixed baseline lengths $\rho=25\,{\rm G}\lambda$ and $\rho=5\,{\rm G}\lambda$. The longer baseline consistently samples a greater variation in the phase.
}
    \label{fig:onering_orbitslice}
\end{figure*}

In the single-ring case, any combination of stretching and shifting preserves the discontinuous phase jumps between $\psi = 0$ and $\psi = \pi$ across a baseline length of approximately $\frac{1}{2R}$. This agrees with the behavior of the 0$^{\rm th}$-order Bessel function at asymptotically long baselines:
\begin{align}
    V(\rho) \approx  \frac{1}{\pi \sqrt{R\rho}} \cos{(2\pi R \rho - \frac{\pi}{4})}.
\end{align}
Here, $\rho$ is the radial baseline in the Fourier domain. The effects of shifting and stretching factorize in the Fourier domain, and so their phase signatures simply add. \autoref{fig:onering_orbitslice} displays how shifting the center of the ring introduces a phase gradient which generally creates the most prominent and unambiguous effect in the phase response, especially on long baselines.
 A greater shift gives a steeper gradient, and thus a more rapidly varying phase. 
 For instance, for a $1\,\mu {\rm as}$ shift, a $25\,{\rm G}\lambda$ baseline would see a $44^\circ$ phase difference relative to the unshifted phase response. This phase signature persists in a relative sense even as an additional ring is added to the system.

\subsection{Two Rings}
\label{sec:tworings}

\begin{figure*}[ht]
    \centering
    \includegraphics[width=0.7\textwidth]{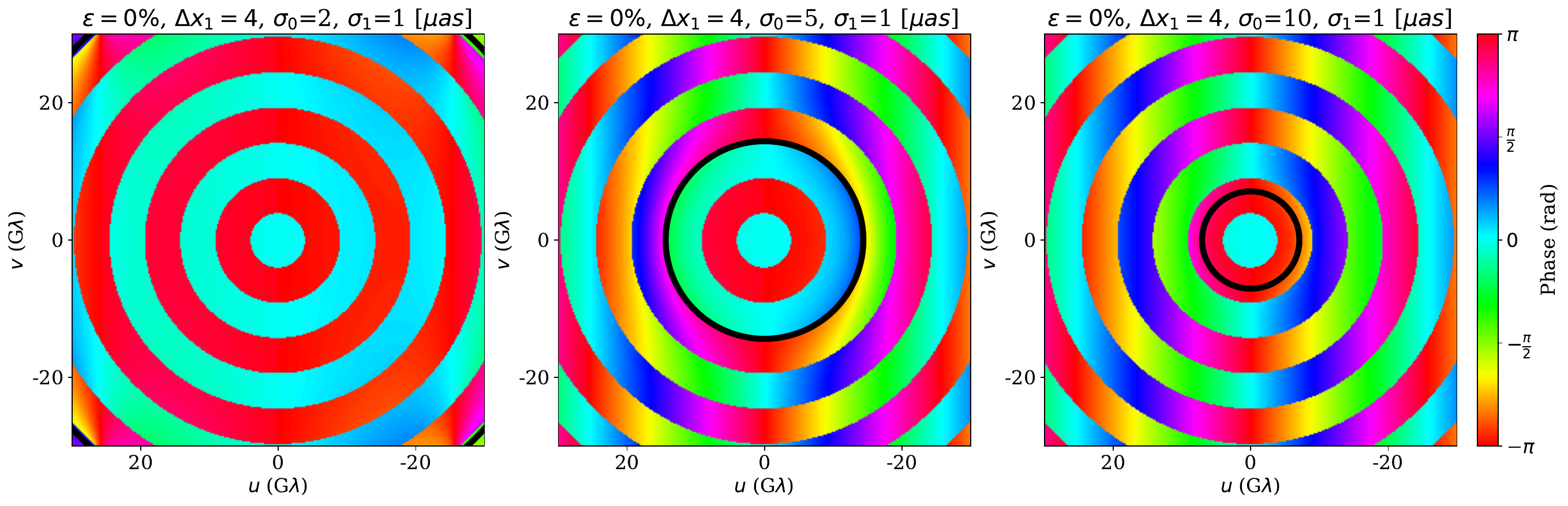}
    \includegraphics[width=0.7\textwidth]{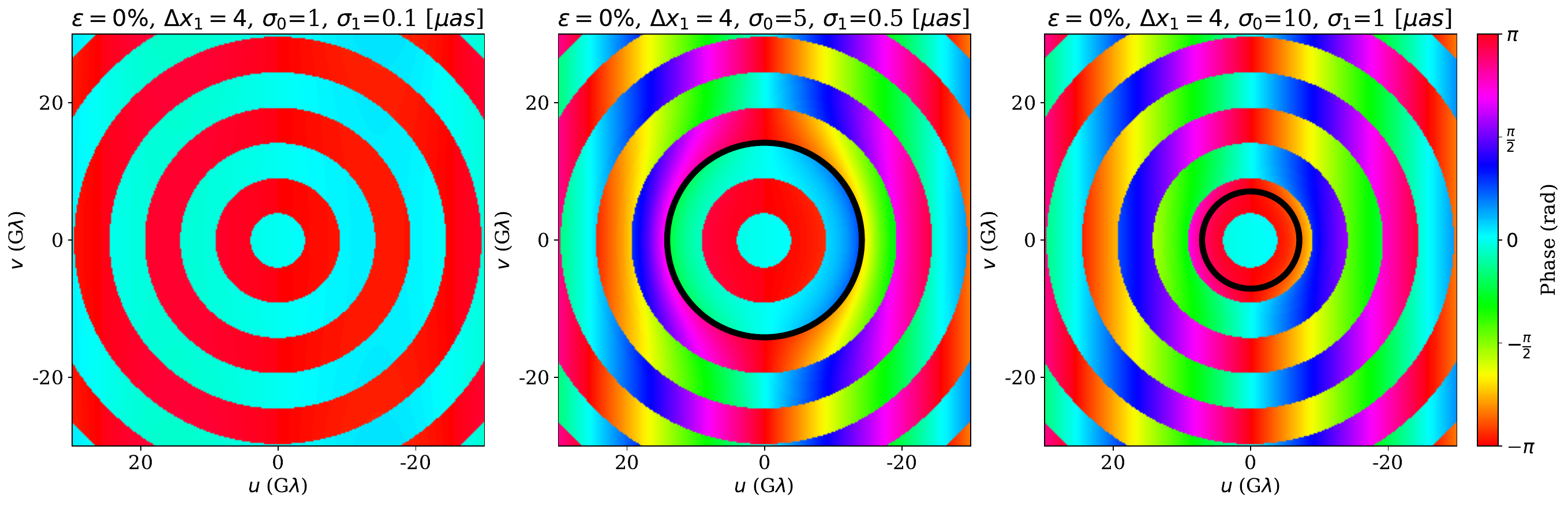}
    \includegraphics[width=0.7\textwidth]{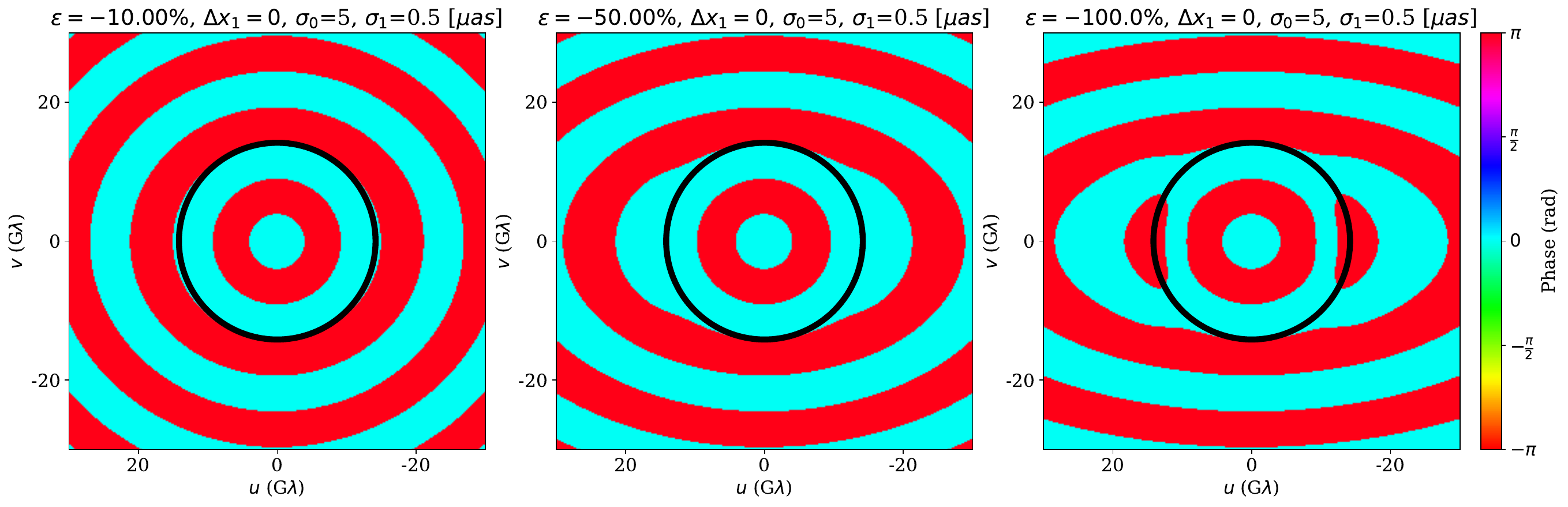}
    \includegraphics[width=0.7\textwidth]{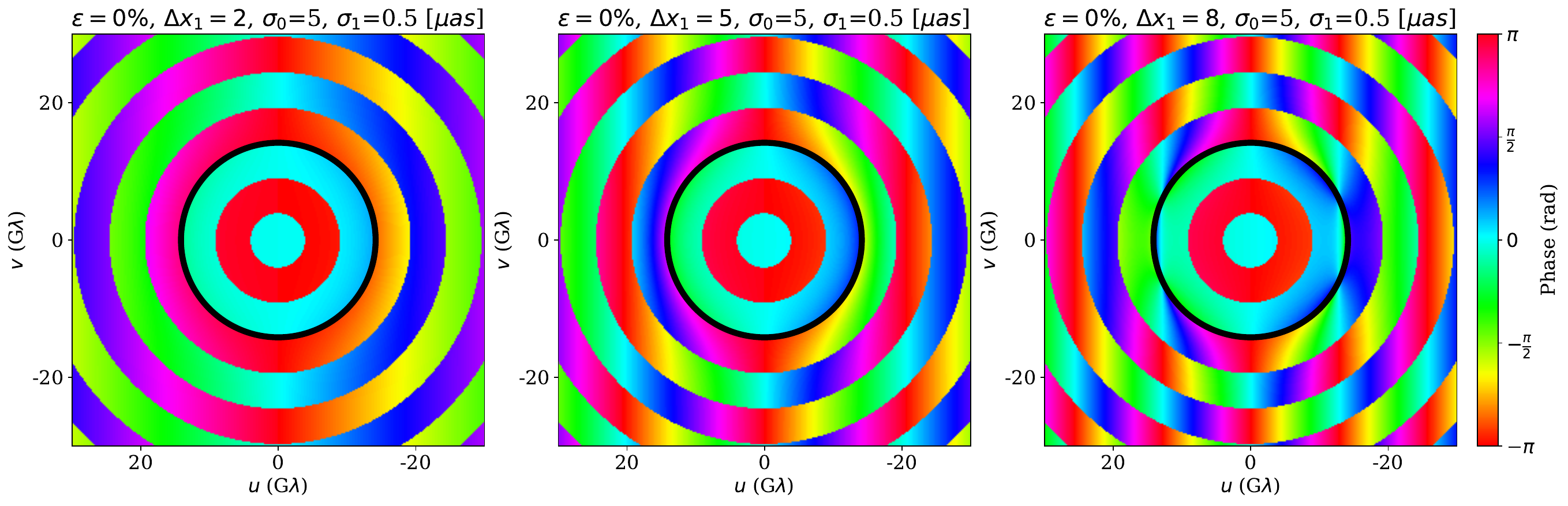}
    \includegraphics[width=0.7\textwidth]{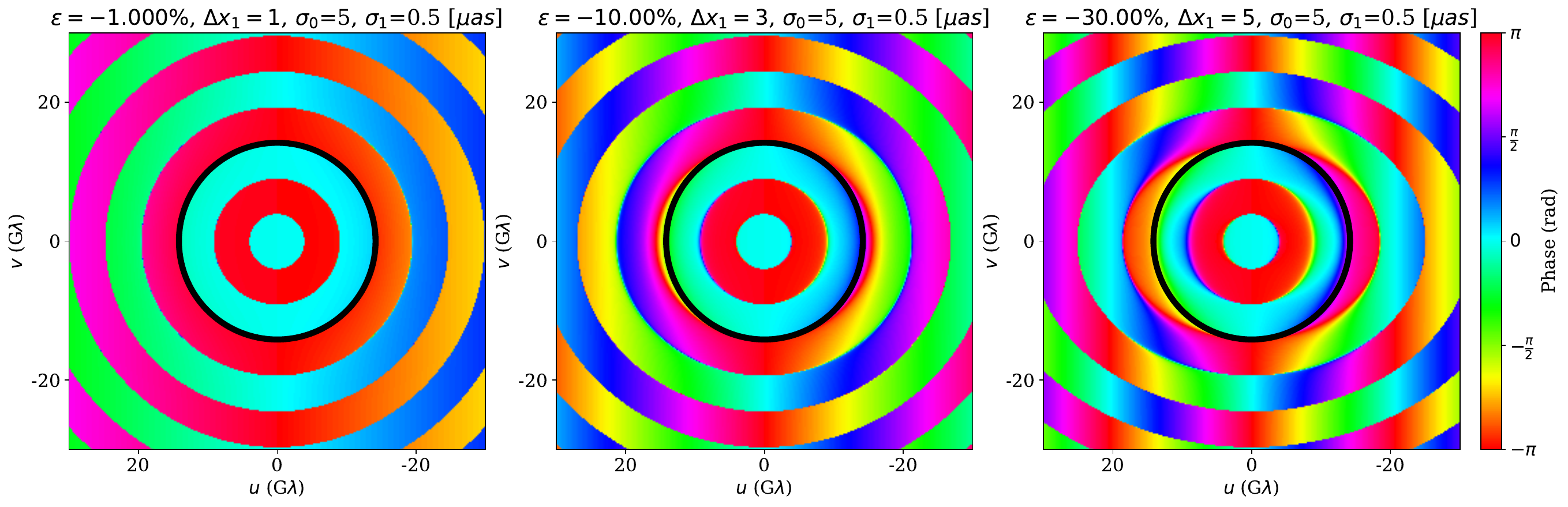}
    
\caption{Interferometric phase response of two superimposed rings, holding the $n=0$ ring centered at the origin in the image domain. Solid black lines show the predicted transition radius between $n=0$ and $n=1$ domination of the phase signature from \autoref{eq:transition}. Row 1: we vary their thickness ratio $\sigma_0/\sigma_1$, with the $n=1$ ring shifted by $\Delta x_1 = 4 \,\mu {\rm as}$ to the right. Row 2: we scale their absolute thickness while holding the thickness ratio fixed at $\sigma_0/\sigma_1=10$, with the $n=1$ similarly displaced. Row 3: we vary the degree of stretching $\epsilon$ for the $n=1$ ring, fixing its absolute thickness at $\sigma_1= 0.5\, \mu {\rm as}$ and without displacing its center. Row 4: we vary the displacement $\Delta x_1$ of the $n=1$, shifting its center to the right. Row 5: we vary both the shift and stretch of the $n=1$ ring. The phase slope along each annular slice grows linearly with the baseline length, becoming more visible at longer baselines. Rows 1-4 preserve the discontinuous phase jumps of the Bessel function. In the last row, however, the mix of the two distortions blurs the discontinuity and allows for a continuous phase slope across slices. The phase jumps are restored at longer baselines.} 
    \label{fig:tworing_gallery}
\end{figure*}

In practice only the relative position of image features (such as a pair of rings) is typically observationally accessible using VLBI, as discussed briefly in \autoref{sec:intro}. 
To isolate the signatures of the $n=1$ ring in the presence of a more weakly spin-distorted $n=0$ ring, we fix the center of the $n=0$ ring at the origin and leave it undistorted, instead focusing solely on distortions to the $n=1$ ring. 
By allowing each ring to take on a distinct flux and thickness, we obtain for their combined visibility response:
\begin{align}
\label{eq:tworingsvis}
    V(u,v) &= F_0 \, e^{-2\pi^2 \sigma_0^2 (u^2+v^2)} J_0\left(2 \pi R_0 \sqrt{u^2+v^2}\right) \\
    \nonumber  & \hspace{-1cm} {} + F_1 \, e^{-2\pi^2 \sigma_1^2 (A_1^2u^2+B_1^2v^2)} J_0\left(2 \pi R_1 \sqrt{A_1^2u^2+B_1^2v^2}\right) \\
    \nonumber  & \hspace{3.3cm} {} \times e^{-i2\pi(u\Delta x_1 + v\Delta y_1)}. 
\end{align}
The parameters $F_i$, $\sigma_i$, $R_i$, $A_i$, $B_i$, $\Delta x_i$, $\Delta y_i$ are the flux, thickness, angular radius, horizontal and vertical stretch, and horizontal and vertical shift of the $n=i$ ring respectively.  
\autoref{fig:tworing_gallery} displays the rich landscape of visibility phase responses generated by varying the aforementioned parameters.

\begin{figure*}[ht]
\centering      
    \includegraphics[width=0.9\textwidth]{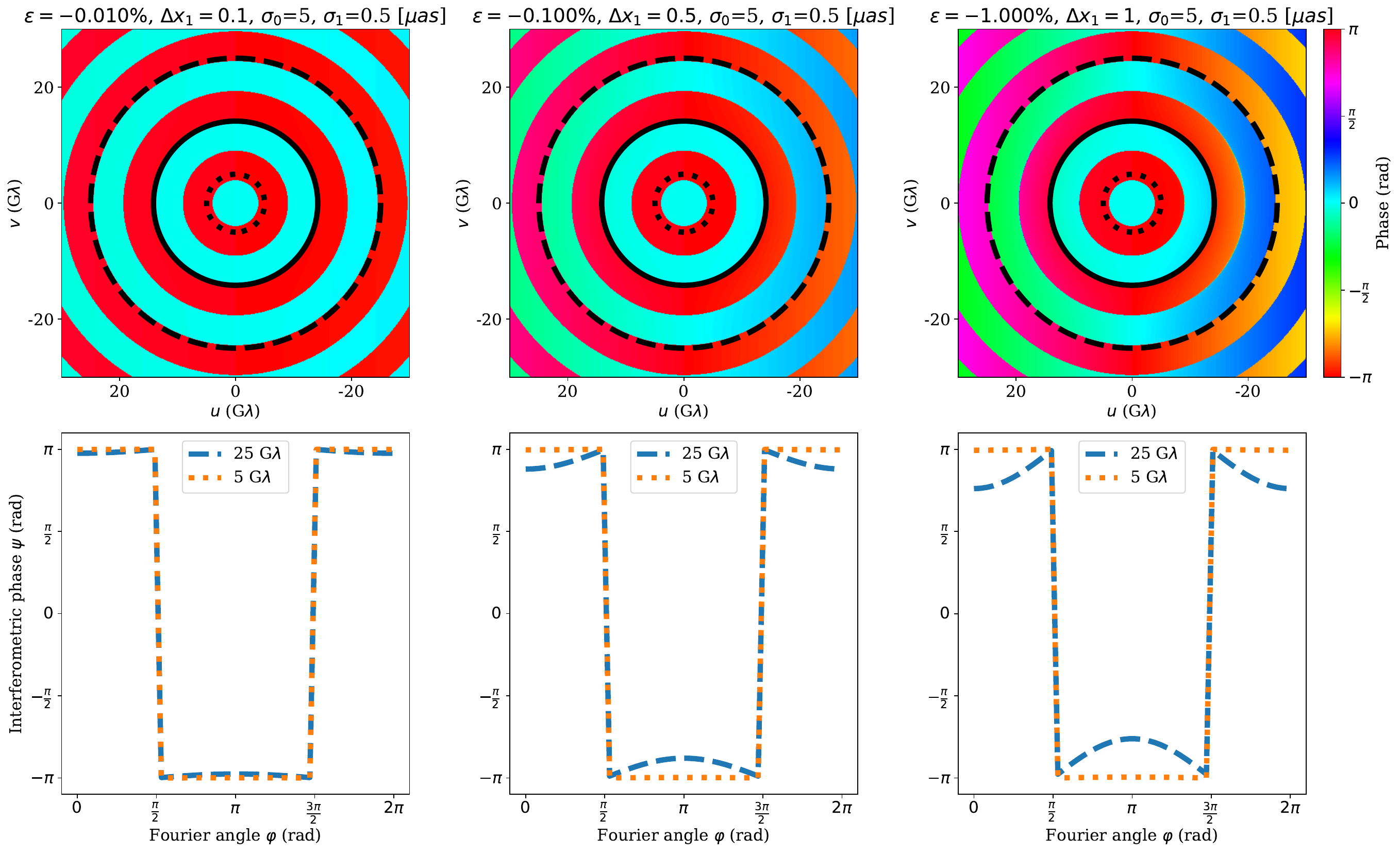}
 
\caption{ Same as \autoref{fig:onering_orbitslice} but for models with two rings: one with $\sigma_0=5\,\mu{\rm as}$, and a thinner ring with $\sigma_1=0.5\,\mu{\rm as}$. As in \autoref{fig:tworing_gallery}, solid lines show the predicted transition radius between $n=0$ and $n=1$ domination of the phase signature. Since the ring thickness and flux are uniform across all three panels, $\rho_T$ is constant as well.
The left panel corresponds to the case where the rings are nearly overlapping and the thinner ring is nearly circular ($\Delta x_1 = 0.1 \,\mu{\rm as}$, $\epsilon=-10^{-4}$). The middle panel has a larger displacement between the rings, and the thinner ring is slightly more stretched ($\Delta x_1 = 0.5 \,\mu{\rm as}$, $\epsilon=-10^{-3}$). The right panel has the greatest displacement and stretching of the thin ring ($\Delta x_1 = 1 \,\mu{\rm as}$, $\epsilon=-10^{-2}$).
}
\label{fig:tworing_orbitslice}
\end{figure*}

While we explore a wide variety of parameter combinations, we generally expect the direct image flux and thickness to greatly exceed those of the $n=1$, with at least $\sigma_0 \gtrsim 10 \sigma_1$ and $F_0 \gtrsim 5 F_1$ in ray-traced simulations \citep[see, e.g.,][]{Johnson_2020, Tamar_2024}. Therefore we vary the geometry of the two-ring system while fixing $F_0=1$ and $F_1=0.1$. Row 2 of \autoref{fig:tworing_gallery} shows how the radius for the transition to photon ring-dominated behavior shifts as the thicknesses are changed while maintaining their ratio, following the transition radius in the $(u,v)$ plane, $\rho_T$, described for a pair of aligned Gaussian rings in \citet{Tamar_2024}:
\begin{align}
    \rho_T &= \sqrt{\frac{\ln\left(F_0/F_1\right)}{2\pi^2 \left(\sigma_0^2 - \sigma_1^2\right)}}.
    \label{eq:transition}
\end{align}
We indicate this radius with a solid black line in each panel of \autoref{fig:tworing_gallery} and \autoref{fig:tworing_orbitslice}.
As is apparent in the top left plot of \autoref{fig:tworing_gallery}, baselines shorter than this radius do not exhibit strong phase slope structure. Given that we are most interested in the relative astrometry of the $n=1$ and $n=0$ rings, baselines sampling both sides of this transition are needed to robustly measure the shift between sub-images. As we will find, this corresponds to measurements on both Earth-Earth and Earth-Space baselines, which together will naturally represent the relative displacement in phases on closure triangles with two Earth observatories and one space observatory.

In general, we expect the brightness of each ring to be comparable in the optically thin limit, so the thickness ratio of the rings is comparable to the flux ratio. Thus, we choose $\sigma_0 = 5 \, \mu {\rm as}$ and $\sigma_1 = 0.5 \, \mu {\rm as}$ unless otherwise noted. Fixing the thickness and flux allows us to isolate the effects of stretching and shifting of the $n=1$ ring. In Row 3 of \autoref{fig:tworing_gallery} we consider only vertical stretching of the $n=1$ ranging from $\epsilon=-10.00\%$ to $\epsilon=-100.0\%$. In Row 4 we do not stretch, but only shift the thinner ring by a distance ranging from $2 \, \mu {\rm as}$ to $8 \, \mu {\rm as}$. Row 5 shows the combined response of a stretch ranging from $\epsilon=-1.000\%$ to $\epsilon=-30.00\%$, and a shift ranging from $1 \, \mu {\rm as}$ to $5 \, \mu {\rm as}$.

As in the case of the single ring, stretching the $n=1$ keeps the visibility real-valued. At the transition radius, the asymmetric response of the stretched $n=1$ begins to dominate over the symmetric response of the $n=0$, and the visibility exhibits non-circular phase flips.
Shifting introduces along each annular slice of the visibility a continuous phase gradient with a steeper slope at longer baselines, consistent with \autoref{eq:tworingsvis}. Separately shifting or stretching the $n=1$ preserves the discontinuous phase jumps between slices, as in the case of a single ring. Unlike for a single ring, however, the combination of stretching and shifting can indeed blur these discontinuities, producing a continuous phase slope along a sector of baselines at length scales dominated by the $n=0$.

Looking back at \autoref{bhspin}, we now have the necessary understanding to explain the phase responses of the low and high spin values. At low spin the shift and stretch are hardly perceptible and thus the visibility closely matches the real-valued Bessel function of zeroth order with its discontinuous phase flips. 
At high spin, the shift of the photon ring induces a phase slope, which becomes more pronounced at longer baselines due to its linear dependence on the latter. Moreover, the subtle stretching of the $n=1$ ring blurs the phase flips into a continuous gradient with a slight spiral quality.

For a mass and inclination similar to those expected for M87* \citep[see, e.g.,][]{Mertens_2016, PaperVI}, a maximally spinning black hole would induce a warping of approximately $3\%$ in the critical curve, while it would produce a shift of the critical curve from the Bardeen coordinate origin of approximately $2\, \mu {\rm as}$ \citep{Johannsen_2010}. Though the $n=1$ ring does not perfectly track the critical curve, this leads us to expect that the most significant and visible effect of spin on the phase signature will be the phase slope introduced by shifting the $n=1$ off-center. 
In \autoref{fig:tworing_orbitslice}, we examine the phase signature on terrestrial and space-VLBI baselines for a system with two rings, with the thinner ring tuned to match the properties in \autoref{fig:onering_orbitslice}. 

If there were no distortions or displacement of the rings, the phase sampled along any fixed baseline length with varying direction would remain constant at either $\psi=0$ or $\psi=\pi$. However, the space-VLBI baseline at 25\,G$\lambda$ samples a phase that fluctuates along the baseline angle; these fluctuations grow with the distortion of the thinner ring, even in the presence of an undistorted bright, thick ring. We now turn to inference-based tests of the utility of both interferometric amplitude and phase for constraining these morphological differences.

\section{Morphological Constraints from Amplitude and Phase}
\label{sec:observability}

We have shown that stretching the $n=1$ ring shifts the phase discontinuities of the ring response; however, as has been well-described in \citet{Johnson_2020}, \citet{Gralla_2020_shape}, and elsewhere, changes in the shape of the $n=1$ ring manifest continuously in the amplitude. So, one might expect that a sparse, discrete sampling of the $(u,v)$ plane would better constrain the $n=1$ asymmetry with amplitudes rather than phases, while phases more naturally capture translation.

\begin{figure*}[ht]
    \centering
    \includegraphics[width=\textwidth]{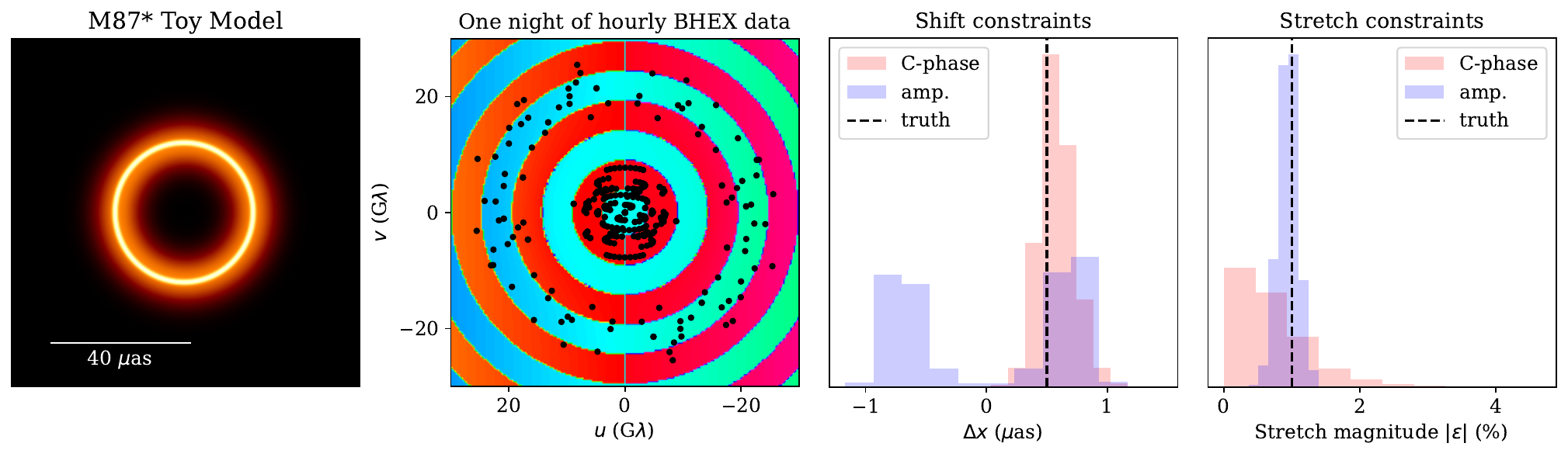}
    \caption{Example measurements of the spin axis-perpendicular relative displacement of the direct and indirect image, as well as the stretch magnitude of the indirect image, fit to a single night of synthetic closure phase or amplitude data with a ground array co-observing with BHEX. First panel: underlying toy model of the two-ring system in M87*. Second panel: a synthetic data set from BHEX overlaid with the underlying model phases. Third panel: horizontal shift constraints with closure phase and amplitude. Fourth panel: same as the third panel for stretch magnitude. 
    }
    \label{fig:datafit}
\end{figure*}

We investigate this difference using mock BHEX observations in \autoref{fig:datafit}, observing a two-ring model with $F_0=1$ Jy, $F_1=0.1$ Jy, $\sigma_0=5\,\mu {\rm as}$, $\sigma_1=0.5\,\mu {\rm as}$, $\Delta x=0.5\,\mu {\rm as}$ (we drop the index as the $n=0$ remains centered), and $|\epsilon|=1\%$. We fix both ring radii to $20\,\mu {\rm as}$. We use the code \texttt{ngEHTsim} to simulate a night of synthetic BHEX data, emulating frequency phase transfer between an 80 GHz low band and a 240 GHz high band \citep{RD_ngeht, Pesce_2024}, accounting for atmospheric effects in its weather model. We assume ground participation from the Atacama Large Millimeter Array (high band), the Green Bank Telescope (low band), the Greenland Telescope (high band), the IRAM 30 meter telescope (low and high bands), the Submillimeter Array (high band), the James Clark Maxwell Telescope (low and high band), the Kitt Peak Observatory (high band), and the Large Millimeter Telescope (low and high band). We assume that BHEX itself lies in a circular polar orbit with a 12-hour period and right ascension of the ascending node oriented 90 degrees from M87*, so that the orbit appears nearly circular from M87*. The synthetic data specifications of dish sensitivity are broadly consistent with the projected performance outlined in recent BHEX literature \citep{Tong_2024, Marrone_2024, Lupsasca_2024}.

Interferometers such as the EHT typically rely on the ``closure phase" technique to eliminate unknown gain contributions to the phase at each observatory \citep{TMS}. Closure phases are formed by summing the interferometric phase around closed triangles of baselines. These quantities gain calibration invariance at the penalty of sacrificing some information; in particular, closure phases are invariant under image translation. However, closure phase can still provide information about relative displacement between multiple image features. In the context of measuring the displacement of the photon ring, it then becomes crucial to observe two subrings simultaneously, to discern their relative separation. Notably, high-$n$ rings closely approximate the critical curve, so the largest relative effects are expected between $n=0$ and $n=1$.

As shown in the second panel of \autoref{fig:datafit}, the synthetic data sample a range of visibilities at which the phase slope caused by the modest separation between the rings is prominent. We then fit the two-ring model to the synthetic data with either exclusively visibility amplitude information, or exclusively closure phase information using \texttt{eht-imaging}'s wrapper around \texttt{Dynesty}, a dynamical nested sampling code \citep{Chael_2016, Speagle_2020}. These fits use Gaussian likelihoods derived from the synthetic thermal noise on each baseline measurement to sample the preferred model parameters for the flux, diameter, stretch, and relative position of the two rings. The particular form of the likelihood for the closure phase is taken from equation 19 in \citet{Chael_2018}, which developed the closure imaging capabilities in \texttt{eht-imaging}. This likelihood is defined circularly in closure phase space, folding in errors on individual visibilities through propagation into the error on the complex bispectrum, and thus its phase. Near nulls in the amplitude response, the signal-to-noise ratio falls, causing large, non-Gaussian phase errors which implicitly downweight contributions to the likelihood of the closure phase. As a result, the primary contributor to the closure phase likelihood are points far from nulls.

The only difference in modeling choices between the amplitude and closure phase fits is that the total flux of the combined model is fixed when fitting closure phase. We do so because the phase is invariant to a global flux rescaling, so we model the ring fluxes by defining the $n=1$ ring flux relative to the $n=0$ with unit flux. We note that in both cases, the location of the $n=0$ ring can (and indeed must) be pinned to the origin without loss of generality, as only absolute interferometric phase would determine the location of the two rings on the sky, and neither amplitude nor closure phase maintain this information.

The third panel shows that the closure phase, which is insensitive to the absolute positional information of the source, tightly constrains the relative displacement of the rings, while the amplitude is both weaker and direction-degenerate. However, in the fourth panel, the stretching of the ring is much more tightly constrained by the amplitudes, suggesting that the continuous representation of stretching in the amplitude domain more naturally constrains the magnitude of the effect.

We conclude that, for a perfectly specified model with a 0.1 Jy $n=1$ ring, BHEX is capable of sub-$\mu {\rm as}$ relative astrometry and is sensitive to percent-level asymmetries in ring structure. We move on now to a brief examination of the expected magnitudes of these effects in more realistic models of the accretion disk.

\section{Non-spacetime signatures and their visibility response}
\label{sec:nonspacetime}

\begin{figure*}
   \begin{centering}
    \includegraphics[width=0.99\textwidth]{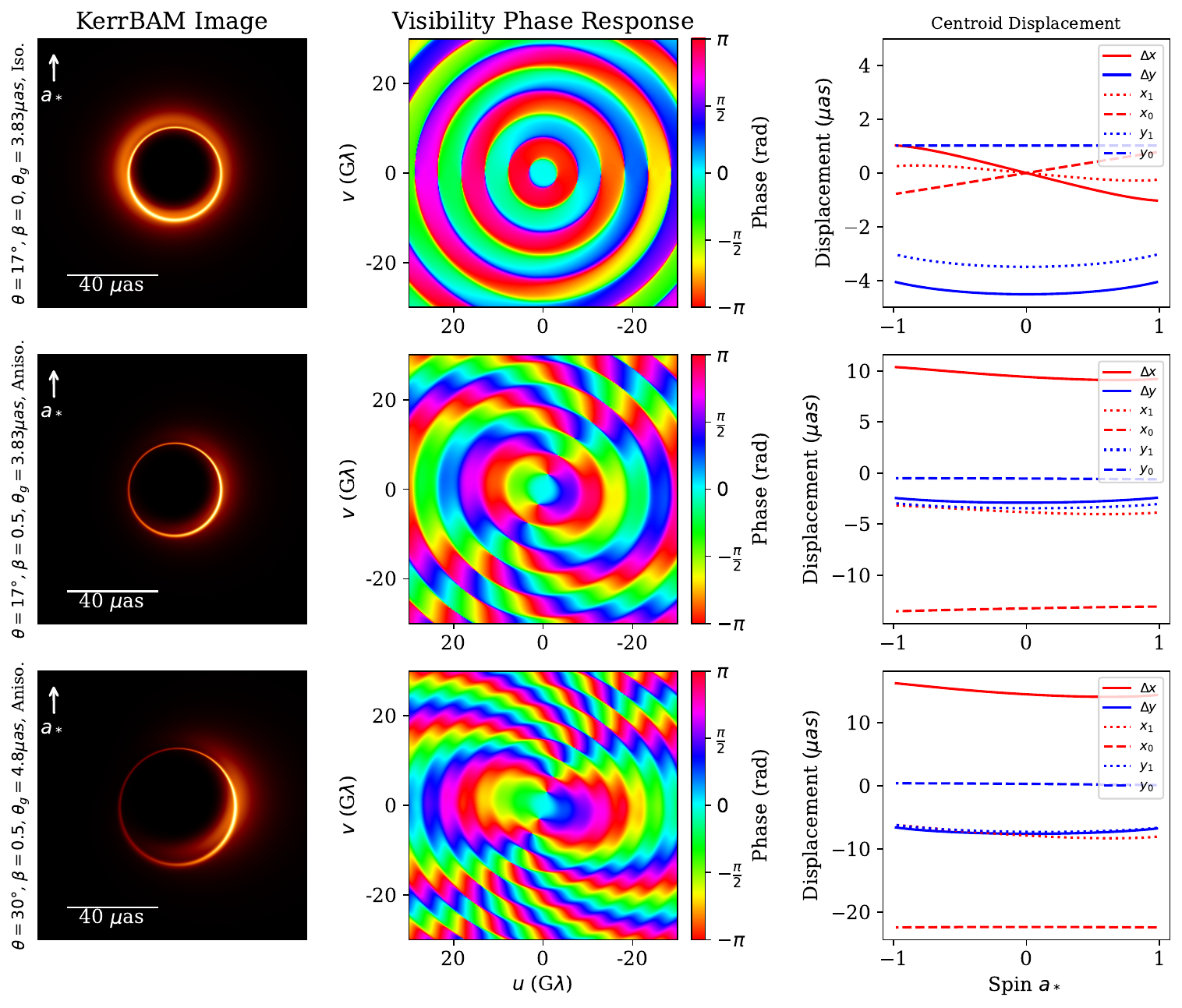}
    \end{centering}
    \caption{
   Semi-analytic accretion disk models along with their visibility phase response and subimage relative displacement as a function of spin.
   First column: three \texttt{KerrBAM}-generated black hole images all with spin $a_* = 0.5$, and varied physical parameters in each row. 
   Labels on the left column indicate the viewer inclination $\theta$, plasma speed $\beta$, angular gravitational radius $\theta_g$, and the isotropic or anisotropic emissivity used.
   Second column: visibility phase in the $(u,v)$ plane corresponding to each image, sampled after centroiding the image. Third column: horizontal (spin-perpendicular) and vertical (spin-parallel) relative displacements, defined as $\Delta x=x_1-x_0$ and $\Delta y=y_1-y_0$ respectively, between the 
   $n=1$ and $n=0$ centroids, as a function of varying spin $a_* \in [-0.98,0.98]$. We also plot absolute horizontal and vertical positions of the $n=0$ and $n=1$ centroids, ($x_n$, $y_n$). While the relative displacement between centroids varies across astrophysical scenarios, its variation with spin is qualitatively similar in all three cases. 
   }
    \label{kerr}
\end{figure*}

We have shown that the relative shift between images expected from changing the spin of a black hole has a pronounced effect in visibilities measured on scales relevant to BHEX. We now investigate whether these simple phase signatures persist in the presence of other sources of image structure and asymmetry, such as accretion disk structures.

One important factor to consider is the effects of magnetic field lines in the emitting plasma; these give angle-dependent emissivity in the rest frame of the emission that can lead to image anisotropies. We also consider plasma speed, which causes asymmetry through Doppler beaming. Additionally, we consider viewer inclination angles beyond face-on, producing non-circular ring shapes and stronger Doppler effects.

To do so, we use semi-analytic KerrBAM models of black hole images to study the visibilities of black holes with varying isotropic (``Iso.'') or anisotropic (``Aniso.'') emission, plasma speed as a fraction of the speed of light ($\beta$), mass-to-distance ratio ($\theta_g$), and angle of inclination relative to our line of sight ($\theta$). Our results are shown in \autoref{kerr}. We take the spin axis to be out of the page for a spin value $a_* > 0$ and into the page for $a_* < 0$. 
We plot the corresponding phase response of each model, and the relative horizontal and vertical offsets, defined as $\Delta x = x_1-x_0$ and $\Delta y=y_1-y_0$ respectively, between the $n=1$ and $n=0$ centroids as a function of spin. 
In addition to the relative shift, the third column of \autoref{kerr} tracks the absolute positions of the $n=0$ and $n=1$ centroids.

Since tracking the positions of geometric centers of the two rings is not simple in the presence of non-uniform brightness, we instead use their centers-of-light (or ``centroids''). For an isotropically emitting, rotationally symmetric accretion disk around a Kerr black hole viewed face-on, the position of the geometric center is equivalent to the position of the centroid. Thus we use the displacement of the $n=1$ centroid to approximate the actual ring displacement due to spin.

Naively, we might expect the shift of the $n=1$ geometric center due solely to spin should lie only on the $x$-axis, perpendicular to spin, and not on the $y$-axis. However, the introduction of new sources of image asymmetry complicates this simple trend, in part due to the usage of the centroid.

In \autoref{kerr} we show that as a function of spin the relative centroid displacement of the $n=1$ ring exhibits a consistent trend across various changing physical parameters, and the phase responses resemble those of the geometric two-ring models, albeit with additional phase ringing arising from added image asymmetries.
In the first column we compare the images of the rings of three different black holes with fixed spin $a_*$, varying $\theta$, $\beta$, $\theta_g$, and the presence of synchrotron emission.
The second column displays visibility phase responses of the three black holes respectively. Consistent with our discussion in \autoref{sec:intro} is the amplification of the stretching effect with a higher $\theta$ in the third row. Interestingly, the low-inclination model closely resembles the two-ring models from \autoref{sec:tworings}, although \autoref{kerr} uses semi-analytic ray-tracing and includes various plasma and position parameters, while \autoref{fig:tworing_gallery} uses idealized analytic Gaussian-convolved delta rings with no physics. 

In particular, introducing additional physical effects does not seem to change the overall behavior of the $n=1$ centroid as a function of spin, as evident from Column 3 of \autoref{kerr}. Though the precise values of the shift vary among the three black holes, they all exhibit a steady downward slope in the $x$-coordinate.
The top plot, corresponding to the symmetric isotropic black hole with low inclination ($\theta = 17^\circ$), shows how the relative displacement $\Delta x$ varies from approximately $1\, \mu {\rm as}$ to $-1\,\mu {\rm as}$ as spin goes from $-0.98$ to $0.98$. 
 Introducing a nonzero $\beta$ and synchrotron radiation translates $\Delta x$ by approximately $10 \, \mu {\rm as}$, as shown in the second plot.   
 The third and final black hole is larger and has a greater viewing inclination, so its relative centroid shift is skewed to around $15 \, \mu {\rm as}$, and varies over a slightly wider interval of approximately $3 \, \mu {\rm as}$. 
For all three examples, the $y$-coordinate of the relative displacement remains mostly steady, exhibiting only a subtle shift that is symmetric across negative and positive spin values. 
It is perhaps surprising that drastically varying the surrounding physics of the black holes amplifies but does not break the trend in photon ring displacement. Nonetheless it is comforting to see that the most pronounced effect of spin is in fact persistent across different physical scenarios.

Note that while we expect the geometric centers of the direct and indirect image to shift in the same direction with spin \citep{Gates_2026}, the behavior of their respective brightness-weighted image centroids depends additionally on the brightness distribution of the ring, which is particularly sensitive to frame-dragging effects from the spinning black hole. For spin directed north in the image, the location of images shifts west (to the right), while the apparent Doppler beaming added by frame dragging causes a brightness asymmetry favoring the east (on the left). As a result, frame-dragging and geometric shift push the centroid in opposite directions. The dominant effect is tuned by spacetime parameters, including the viewing inclination, as well as astrophysical parameters that tune the on-sky area of each image.

Consider the black hole in the top row of Figure 7, in which the surrounding plasma is stationary in the ZAMO frame and emits isotropic emission. The position of the direct image centroid is dominated by frame-dragging and thus $x_0$ exhibits an overall shift to the left as spin increases, as shown by its positive slope. Meanwhile the indirect image centroid is dominated by the amplified geometric shift of the ring, so $x_1$ exhibits an overall shift towards the right.

The middle row of \autoref{kerr} features a black hole with non-zero plasma speed and anisotropic synchrotron emission. The tension between frame-dragging and geometric shift persists even in this more complicated astrophysical scenario. While both the $n=0$ and $n=1$ centroids lie to the right across all values of spin, the slope of $x_0$ with respect to spin is again positive indicating a leftward shift, while the slope of the $x_1$ is negative indicating a rightward shift.

The bottom row of \autoref{kerr} features a larger black hole, viewed at greater inclination, with the same plasma speed and anisotropic synchrotron emission profile as the middle row.
$x_0$ now follows a nearly flat downwards parabolic trajectory with a negligible variation with spin. Yet the slope of the $x_1$ remains approximately linear and negative, much like in the previous rows. These results further indicate that while the behavior of the direct image is sensitive to both the surrounding astrophysics and frame-dragging, the indirect image is largely dominated by the spacetime geometry near the horizon, in particular by the shift due to spin. Much like that of the indirect image, the relative displacement $\Delta x$ between the two centroids remains approximately linear and negative across all three scenarios.

These models show that, even when astrophysical effects contribute to the absolute centroid displacement, the relative translational effect of spin is persistent across scenarios and nearly linear, with a magnitude $\Delta x_{\rm spin}$ of approximately $1\, \mu{\rm as}$ per unit spin for M87*. A translation $\Delta x$ causes a phase signature on a baseline of length $\rho$ aligned with the translation of $\Delta\psi = 2\pi\rho\Delta x $.  In the high signal-to-noise ratio (SNR) regime, the phase error on a baseline is simply $\sigma_\psi \approx 1/S$, where $S$ is the SNR. We may thereby define an astrometric resolution $\theta_{\rm ast}$ by equating the phase error to the phase signature from the translation (equivalently a $1\sigma$ marginal detection of the shift):
\begin{align}
    \theta_{\rm ast} \equiv \frac{\sigma_\psi}{2 \pi \rho}.
\end{align}
In order to determine the detectability of the spin signature, the angular shift rate can be related to the astrometric resolution at the longest baseline $\rho_{\rm max}$:
\begin{align}
    \Delta x_{\rm spin} 
    &\approx \theta_{\rm ast},\nonumber\\
    &\approx \frac{1}{2\pi S \rho_{\rm max}}. \\
    \rightarrow S_{\rm min} &\gtrsim \frac{1}{2\pi \rho_{\rm max} \Delta x_{\rm spin}}
\end{align}
Thus, in order to measure the M87* spin at the 10\% level on a 25 G$\lambda$ baseline, the requisite SNR is given by substituting 10\% of the modeled $\Delta x_{\rm spin}$ for M87*:
\begin{align}
    S &\gtrsim \frac{1}{2\pi \times 25{\rm G}\lambda \times 0.1 \times 1\mu {\rm as}},\nonumber\\
    &\gtrsim 13.
    \label{eq:snr}
\end{align}
That is, a typical BHEX baseline must be able to integrate down to an SNR of 13 in order to measure the displacement phase effect from a 0.1 difference in spin on that baseline, under conditions of otherwise knowing the plasma properties and their effect on the subimage centroids. Since the displacement effect is present most cleanly in the time-averaged image, this $S$ value should be viewed as a joint requirement for the total number of observations together with the snapshot observational sensitivity. In particular, we expect repeated observations with a more typical SNR of 3 to 5 on individual visibilities to accumulate over observing campaigns. Our SNR requirement can be viewed as a requirement for the combined sensitivity of measurements over a range of long baselines exceeding $\rho_T$, merged temporally or spatially to accrue a sufficient understanding of the relative astrometry. We also note that these sensitivity requirements apply comparably to the ground, but that terrestrial baselines will generally exceed the expected Earth-space baseline SNR due to the larger signal on shorter baselines, and larger telescope diameters on Earth.
While we leave the fitting to synthetic BHEX data of astrophysical models such as the ones in \autoref{kerr} to future work, we note that similar fits to synthetic EHT data show that spin displacement is not adequately constrained without space baselines \citep{Chang_2024, Nowicki2026}.

\section{Conclusion}
\label{sec:conclusion}

We have analyzed the interferometric phase response expected from the image of a black hole accretion disk and its first lensed photon ring, assessing the measurement of the stretching and shifting of images imposed by black hole spin. First, we presented a simplified and fully analytic geometric analysis, modeling the two subimages as Gaussian-convolved ellipses, with physically motivated parameters for their thicknesses, fluxes, and distortions. 
We showed that imposing a relative displacement of the two rings commensurate with expectations from general relativistic predictions gives a phase slope through the Fourier shift theorem that is readily measured on Earth-space baselines.
 
We then used geometric self-fits to synthetic BHEX data to show that even when absolute astrometry is lost, the displacement between the direct and indirect image is still recoverable from interferometric closure phases, which hold relative astrometric information that preserves the spin drift between the $n=0$ and $n=1$ images. This behavior contrasts with the constraints provided by interferometric amplitude, which more naturally constrain the stretch of the ring system while only weakly and degenerately constraining the shift. 

Next, we studied semi-analytic ray-traced models of black holes where we varied physical parameters other than spin, and modeled their effects on the relative centroid displacement of the first two rings. We showed that while the relative displacement of the centroids is in fact heavily dependent on these other properties of the black hole, the additional relative displacement owing to spin is approximately linear and similar in magnitude across models, varying most strongly with inclination. After isolating the spin effect, we found its magnitude to be approximately $1 \, \mu {\rm as}$ per unit spin for an M87*-like black hole system, setting a benchmark for the effective astrometric resolution of interferometers seeking to measure spin from relative displacement.  Future studies may refine the spin displacement effect through brightness-unweighted treatments of subimage locations, downstream from imaging reconstruction and image-domain feature extraction. 

In order to make determinations about spin from astrometric displacement or interferometric phase alone, it is apparent that confounding effects from the accretion disk must be studied and modeled in parallel with the spacetime. The presence of non-spacetime signatures in the phase at a particular baseline means that a simple pipeline involving, say, incoherent averaging of the closure phase on a large Earth-space triangle will not, on its own, produce a robust measurement of spin, instead favoring fits to a time-varying accretion model on top of a static set of black hole parameters. Of particular interest are models which permit turbulent fluctuations in the accretion flow; recent work has shown that simply allowing time variability in the parameters of axisymmetric plasma models is insufficient to capture spacetime parameters in parallel with accretion properties \citep{Chang_2025}. Instead, fitting a long term average for the accretion flow's physical properties alongside the spacetime parameters, while allowing substructure in individual epochs to be drawn from a fitted distribution for fluctuations of the plasma density, temperature, and magnetic field will naturally provide the astrophysical uncertainty on spin that we only begin to estimate in this work. These models are currently under development, and will play a key role in BHEX measurement of black hole properties, beyond astrometry of the photon ring.

More complete model-fitting studies with full treatments of the emission morphology and spacetime properties may not only help us to determine spin more accurately and with greater precision, but also to better characterize yet-poorly-understood accretion properties and construct a better picture of the physics surrounding the black hole. We have shown phase to be a powerful and irreplaceable component of the photon ring measurement problem.
Taken together with the partially disjoint spin constraint provided by interferometric amplitudes, our results suggest that Earth-space interferometry will likely be strongly and jointly constraining of both the spacetime and accretion disk properties, such as magnetic field and fluid velocity orientation.
 
\section{Acknowledgements}
We gratefully acknowledge helpful conversations with Dr. Delilah Gates and Dr. Dominic Chang.

We acknowledge financial support from the National Science Foundation (AST-2307887). S.~G-L. is funded through generous support from Mr.\ Michael Tuteur and Amy Tuteur, MD.  This publication is funded in part by the Gordon and Betty Moore Foundation, Grant GBMF12987. This work was supported by the Black Hole Initiative, which is funded by grants from the John Templeton Foundation (Grant \#62286) and the Gordon and Betty Moore Foundation (Grant GBMF-8273) - although the opinions expressed in this work are those of the author(s) and do not necessarily reflect the views of these Foundations. 

\begin{acknowledgements}
\end{acknowledgements}

\bibliography{bib}{}
\bibliographystyle{aasjournal}

\end{document}